\documentstyle[12pt]{article}
\begin{document} 
\pagestyle{empty}
\vspace*{-40pt} 
\begin{center}
\rightline{FERMILAB--Pub--93/399-A} 
\rightline{SUSSEX-AST 94/2-1}
\rightline{astro-ph/9402021}
\rightline{(February 1994)}
\rightline{submitted to {\it Physical Review D}}

\vspace{.2in}

{\Large \bf SECOND-ORDER RECONSTRUCTION \\ 
\bigskip OF THE INFLATIONARY POTENTIAL}\\

\vspace{.2in} 
{\large Andrew R.  Liddle$^1$ and Michael S.  Turner$^{2,3}$}\\
\vspace{.2in}

{\it $^1$Astronomy Centre, Division of Physics and Astronomy\\ University 
of Sussex, Brighton~~BN1 9QH, UK}\\

\vspace{.1in}

{\it $^2$NASA/Fermilab Astrophysics Center\\ Fermi National Accelerator
Laboratory, Batavia, IL~~60510-0500}\\

\vspace{.1in} {\it $^3$Departments of Physics and of Astronomy \&
Astrophysics\\ Enrico Fermi Institute, The University of Chicago, Chicago,
IL~~60637-1433}\\

\end{center}

\vspace{.3in}

\centerline{\bf ABSTRACT} 
\medskip

\noindent To first order in the
deviation from scale invariance the inflationary potential and its first two
derivatives can be expressed in terms of the spectral indices of the scalar
and tensor perturbations, $n$ and $n_T$, and their contributions to the
variance of the quadrupole CBR temperature
anisotropy, $S$ and $T$.  In addition, there
is a ``consistency relation'' between these quantities:  $n_T=
-{1\over 7}{T\over S}$.  We discuss the overall strategy of
perturbative reconstruction and
derive the second-order expressions for the inflationary potential and its
first two derivatives and the first-order expression for
its third derivative, all in terms of $n$, $n_T$, $S$, $T$, and $dn/d\ln k$.
We also obtain the second-order consistency relation, $n_T
=-{1\over 7}{T\over S}[1+0.11{T\over S} + 0.15 (n-1)]$.
As an example we consider the exponential potential, the only known case
where exact analytic solutions for
the perturbation spectra exist.  We reconstruct the potential
via Taylor expansion (with coefficients calculated at both first and second 
order), and introduce the Pad\'{e} approximant as a greatly
improved alternative.

\vspace{0.3in}

\noindent PACS number(s):  98.80.Cq, 98.70.Vc

\newpage 
\pagestyle{plain} 
\setcounter{page}{1} 
\newpage


\section{Introduction}

In inflationary models quantum fluctuations excited on very small length
scales ($\sim H^{-1}\sim 10^{-23}{\rm cm}$) are stretched to astrophysical
scales ($\sim 10^{25} {\rm cm}$) by the tremendous growth of the scale factor
during inflation ($H$ is the value of Hubble parameter during inflation)
\cite{kt}.  This results in almost scale-invariant spectra of scalar
(density) \cite{scalar} and tensor (gravitational-wave) \cite{tensor} metric
perturbations.  Together with the prediction of a spatially-flat Universe
they provide the means for testing the inflationary paradigm.  The tensor
fluctuations lead to cosmic background radiation (CBR) anisotropy and a
stochastic background of gravitational waves with wavelengths from about
$1\,$km to over $10^4$ Mpc.  The scalar fluctuations also lead to CBR 
anisotropy and seed the formation of structure in the Universe.

The amplitudes and spectral indices of the metric fluctuations can be
expressed in terms of the inflationary potential and its derivatives,
evaluated at the value of the scalar field when astrophysically interesting
scales crossed outside the horizon during inflation (from
galactic scales to the presently observable Universe,
corresponding to the eight e-foldings about 50 e-folds or so
before the end of inflation).  Techniques have been developed
for relating the scalar and tensor spectra to the potential
and its derivatives in an expansion whose small parameter is
the deviation from scale invariance \cite{expand,turner1}.
In particular, the spectral indices and the power spectra of the
fluctuations today can be written as \cite{turner1}\footnote{Several
minor errors in Ref.~\cite{turner1} have been corrected here:
the factor of $H_0^{3+n}$ in Eq.~(A5) should be $H_0^4$;
the factor of $H_0^{3+n}$ in Eq.~(A7) should be $2^{n-1}H_0^4$;
the factor of $1.1(n-1)$ in Eq.~(A8) is more precisely
$1.3(n-1)$; the factor of $1.2n_T$ in Eq.~(A14) is
more precisely $1.4n_T$.}
\begin{equation}\label{eqns}
n = 1 - {x_{50}^2 \over
8\pi} + {m_{{\rm Pl}} x^\prime_{50}\over 4\pi},
\qquad n_T = -{x_{50}^2 \over 8\pi},
\end{equation} 
\begin{equation} 
P(k) = Ak^n|T(k)|^2, \qquad P_T(k) =
A_Tk^{n_T-3} |T_T(k)|^2 \left( {3j_1(k\tau_0 ) \over k\tau_0 } \right)^2,
\end{equation} 
\begin{eqnarray}
\label{scaldef} 
A &=& {1024\pi^3k_{50}^{1-n}\over 75H_0^{4}}\left[ 1+ {7\over 6}n_T + \left(- 
{7\over 3}+\ln 2 +\gamma \right)(n-1)\right] {V_{50}\over m_{{\rm 
Pl}}^4x_{50}^2},\\ 
\label{tensdef}
A_T & = & {8k_{50}^{-n_T}\over 3\pi}\left[ 1 + \left( -{7\over 6} + \ln 2 + 
\gamma \right) n_T \right] {V_{50}\over m_{{\rm Pl}}^4} .  
\end{eqnarray} 
Here $k$ is the comoving wavenumber, $x = m_{{\rm Pl}} V^\prime /V$
measures the steepness of the potential, prime denotes
derivative with respect to the scalar field that drives inflation, subscript
50 indicates that the quantity is to be evaluated 50 e-folds before the end
of inflation,\footnote{The point about which the potential is expanded is in
principle arbitrary.  However, the spectral indices $n$ and $n_T$ can only
plausibly be measured on scales from $1{\rm \,Mpc} -10^4 {\rm \,Mpc}$ and $S$ 
and
$T$ depend upon perturbations on these same scales, so it makes sense to
choose the expansion point to correspond to when these scales
crossed outside the horizon during inflation; in addition,
by taking $k_{50}\tau_0 =1$ several expressions simplify. The precise
number of e-folds before the end of inflation
when these scales crossed outside the horizon
depends logarithmically upon the
energy scale of inflation and the reheat temperature, see
Refs.~\cite{expand,turner1,copeland}; for the sake of definiteness
we take this number to be 50, which can easily
be changed to the correct value for a given inflationary
model.}  $m_{{\rm Pl}} =1.22\times 10^{19}$ GeV is the
Planck mass, $H_0$ is the present value
of the Hubble constant, $\tau_0\simeq 2H_0^{-1}$ is the present conformal age
of the Universe, and $\gamma \simeq 0.577$ is Euler's constant.
Scale-invariant metric perturbations correspond to $(n-1) = n_T =0$.
The functions $T(k)$ and $T_T(k)$
are the transfer functions for scalar \cite{sctf}
and tensor \cite{turner2} metric perturbations
respectively; for $k\tau_0 \ll 100$, both $T(k)$ and $T_T(k)\rightarrow 1$.
The expressions for $n$ and $n_T$ are given to lowest
order in the deviation from scale invariance (hereafter,
referred to as first order), and the expressions for $A$ and
$A_T$ include the lowest-order term as well as the next
correction (hereafter, referred to second order) \cite{SL}.

From these expressions the consequences of the scalar and tensor metric
fluctuations may be computed.  In particular, the contributions to the
variance of the angular power spectrum of the CBR anisotropy on large angular
scales ($l\ll 200$) which arise predominantly due to the Sachs-Wolfe effect
are given by \cite{turner2} 
\begin{eqnarray}\label{eq:sws}
\langle |a_{lm}^S|^2\rangle & =
& {H_0^4 \over 2\pi} \int_0^\infty k^{-2}P(k)|j_l(k\tau_0)|^2dk ,\\ 
\label{eq:swt}\langle |a_{lm}^T|^2\rangle
& = & 36\pi^2 {\Gamma (l+3) \over \Gamma (l-1)}
A_T \int_0^\infty k^{n_T+1}|F_l(k)|^2 |T_T(k)|^2 dk,\\
F_l(k) & = & - \int_{\tau_{\rm
LSS}}^{\tau_0} {j_2(k\tau_0)\over k\tau_0}\left( {j_l(k\tau_0 -k\tau )\over
(k\tau_0 -k\tau )^2} \right) d\tau ,
\end{eqnarray} 
where $\tau_{\rm LSS} \simeq \tau_0/(1+ z_{\rm LSS})^{1/2} \approx
\tau_0/35$ is the conformal age at last scattering ($z_{\rm LSS} \simeq
1100$) and $j_l$ is the spherical Bessel function of order $l$.  (We
note in passing that both expressions are based upon the approximation
that the Universe is matter-dominated at last-scattering; the small
contribution of radiation, about 10\%-20\%, leads to corrections
\cite{turner2} that would have to be included in a
more accurate treatment.  The corrections
to the quadrupole anisotropy are small.)

The contribution of scalar and tensor metric perturbations to the
observer averaged variance of the quadrupole CBR
anisotropy can be computed numerically \cite{turner1}
\begin{eqnarray}
\label{eqS} S \equiv {5 \langle |a_{2m}^S|^2\rangle \over 4\pi} & \simeq &
2.2 \left[ 1 +1.2 n_T + 0.08 (n-1)\right] {V_{50}(k_{50}\tau_0)^{1-n} \over
m_{{\rm Pl}}^4 x_{50}^2},\\ 
\label{eqT} T \equiv {5\langle
|a_{2m}^T|^2\rangle \over 4\pi} & \simeq & 0.61 \left[ 1 +1.4n_T \right]
{V_{50}(k_{50}\tau_0)^{-n_T}\over m_{{\rm Pl}}^4}, 
\end{eqnarray} 
where the
dependence upon $(n-1)$ and $n_T$ is given to first-order.  In evaluating
these expressions the effect of transfer functions is negligible as the
integrals are dominated by $k\tau_0\sim 2$.  For simplicity, following 
footnote 2 we henceforth omit factors of $(k_{50}\tau_0)^{1-n}$ and 
$(k_{50}\tau_0)^{-n_T}$; they are easily re-inserted if needed.

\subsection{First-order reconstruction}

We choose $S$, $T$, $n_T$, and $(n-1)$ as a convenient set of observables; 
other choices are possible and can be
easily transformed to our set.\footnote{For example, in separating the
tensor and scalar contributions to CBR anisotropy one might measure
$l(l+1)\langle |a_{lm}|^2\rangle$ for four values of $l$ (or ranges
centered on four different values of $l$); see e.g.,
Refs.~\cite{CBDES,DKK}.  From these measurements and the known
dependence of $l(l+1)\langle |a_{lm}^S|^2\rangle /4\pi \approx S
(l/2)^{n-1}$ and $l(l+1)\langle |a_{lm}^T|^2\rangle /4\pi
\approx T (l/2)^{n_T}$ upon $S$, $(n-1)$, $T$, and $n_T$,
cf. Eqs.~(\ref{eq:sws}) and (\ref{eq:swt}), our chosen observables
can be extracted.} 
Since $S$, $T$, $n$, and $n_T$ are expressed in terms of the potential and
its first two derivatives, one can invert the expressions to solve for the
potential and its first two derivatives in terms of $S$, $T$, $n$, and $n_T$
plus a ``consistency relation.'' Those expressions are \cite{turner3}
\begin{eqnarray}
V_{50}/m_{{\rm Pl}}^4 & = & 1.65(1-1.4 n_T)T,\nonumber\\
\label{eq:fv} &=&1.65\left( 1+0.20{T\over S} \right) T, \\
V_{50}^{\prime}/m_{{\rm Pl}}^3 & = & \pm 8.3\sqrt{-n_T}T,\nonumber\\
\label{eq:fvp} & = & \pm8.3\sqrt{{1\over 7}{T\over S}} T, \\
V_{50}^{\prime\prime}/m_{{\rm Pl}}^2 & = &
21[(n-1)-3n_T]T,\nonumber\\
\label{eq:fvpp}& = & 21 \left[ (n-1) +0.43{T\over S} \right] T,\\
\label{eq:fcons} n_T & = & - {1\over 7}{T\over S}.
\end{eqnarray} 
In the second expressions for the
potential and its first two derivatives we have used the consistency relation
to express $n_T$ in terms of $T\over S$, as $T\over S$
should be easier to measure \cite{CBDES}.
Note that the sign of $V^\prime$ cannot be determined as it can be changed by a
field redefinition $\phi \rightarrow -\phi$, though a specific choice here 
determines the signs of various later expressions.
This procedure actually generates the full second-order term
for $V_{50}$, while the other expressions are first-order.

In order to actually reconstruct the inflationary potential over the
eight or so e-folds relevant for astrophysics from its value and
first two derivatives one needs to relate the number of e-foldings
from the end of inflation, $N$ (where $dN/dt =-H$),
to the value of the scalar field.  To lowest order the
equation for $d\phi /dN$ follows from the slow-roll equation
for the evolution of $\phi$ and is given by
\begin{equation}
\frac{d\phi}{dN} \simeq \frac{m_{{\rm Pl}}^2}{8\pi} \, \frac{V'}{V},
\end{equation}
where to lowest order the right-hand side is just $\sqrt{-n_T}\,m_{\rm 
Pl}/\sqrt{8\pi}$.

In the next section we discuss the overall strategy of
perturbative reconstruction, and in the following section
go on to derive the second-order expressions for the potential
and its first two derivatives, for the equation relating
$\phi$ and $N$, and for the consistency relation, as well as
the first-order expression for the third derivative.
We finish with a brief discussion of our results.

\section{Perturbative Reconstruction Strategy}

While one can hope to learn about the potential over the interval that
affects astrophysical scales, it is probably not realistic to hope to
learn much about the potential globally without some additional a priori
knowledge (e.g., the functional form of the potential).\footnote{The one
possible exception involves the accurate measurement of the stochastic
background of gravitational waves on scales from $1\,$km to $3000\,$Mpc
(corresponding to $N \simeq 0 -50$) in which case the inflationary
potential could be mapped out directly since the amplitude of the tensor
perturbation on a given scale is related to the value of the potential.}
The fundamental goal of perturbative reconstruction is to use a finite
set of data to reconstruct the inflationary potential over the interval
where the eight or so e-foldings of inflation relevant to astrophysics took
place.  The observational data all trace to the scalar and tensor metric
perturbations, whose observable consequences can be expressed in terms
of the inflationary potential $V$ and its derivatives
$V^{(m)}$ evaluated at some convenient point in this interval.  (For
brevity, in this section we drop the subscript `50' that indicates where the
potential and its derivatives are to be evaluated.)  Once the observables,
e.g., $n_T$, $(n-1)$, $S$, $T$ and so on, have been expressed in terms of the
potential and its derivatives, these expressions can be inverted to
express the potential and its derivatives in terms of the observables,
as well as a consistency relation.  From these the potential can be
recovered by expansion.

In principle, the observables depend upon all the derivatives of the
potential, making the problem appear intractable.  If one is willing to
restrict the problem to flat potentials which lead to nearly
scale-invariant perturbations and nearly exponential inflation, the
problem can be made manageable.  (In the
scale-invariant limit the potential is precisely constant and all its
derivatives vanish.)  In the nearly scale-invariant
limit we have a set, albeit
infinite, of small parameters to expand in: $m_{{\rm Pl}}^m V^{(m)}/V$;
as we shall describe, when calculating to a given accuracy only a small
number of derivatives are needed. Put another way,
the terms involving more derivatives or higher derivatives
are of higher order.  To be more specific, to lowest order
\begin{eqnarray}
T & \sim & {\cal O}(V/m_{\rm Pl}^4), \\
n_T,\  T/S & \sim & {\cal O}[(m_{{\rm Pl}} V^\prime /V)^2],\\
(n-1) & \sim & {\cal O}[(m_{{\rm Pl}} V^\prime /V)^2] +
{\cal O}[m_{{\rm Pl}}^2 V^{\prime\prime}/V];
\end{eqnarray}
that is, only the potential and its first two derivatives come into play.

Corrections from higher derivatives
come into play because of the variation of the potential
and its first two derivatives during the Hubble time or
so that a given scale is crossing outside the horizon
and is becoming a classical metric perturbation.  It is
straightforward to write down the form of the higher derivative
terms expected by using the fact that the variation of a given
derivative over a Hubble due to a higher derivative is:
\begin{equation}
\delta V^{(n)} \sim V^{(m)}\delta \phi^{m-n} \sim
(m_{\rm Pl}^2 V^\prime /V)^{m-n} V^{(m)},
\end{equation}
where the final expression follows by using $d\phi /dN \sim
m_{\rm Pl}^2V^\prime /V$.  The form of the higher-order
terms in the expansions of $T$, $n_T$,
$T/S$, and $n$ can now be written down directly:
\begin{eqnarray}
T & \sim & {\cal O}(V/m_{\rm Pl}^4) \left[ 1+ {\cal O}[(m_{\rm Pl}V^\prime
/V)^2] + {\cal O}[(m_{\rm Pl}^2V^{\prime\prime}/V)(m_{\rm Pl}V^\prime
/V)^2] + \cdots \right] ,\\
n_T,\  T/S & \sim & {\cal O}[(m_{\rm Pl}V^\prime /V)^2]
+ {\cal O}[(m_{\rm Pl}^2V^{\prime\prime}/V)(m_{\rm Pl}V^\prime
/V)^2] \nonumber\\
& + & {\cal O}[(m_{\rm Pl}^3V^{(3)}/V)(m_{\rm Pl}V^\prime /V)^3]
+ \cdots ,\\
n-1 & \sim & {\cal O}[(m_{\rm Pl}V^\prime /V)^2] +{\cal O}[m_{\rm Pl}^2
V^{\prime\prime}/V] + {\cal O}[(m_{\rm Pl}^3V^{(3)}/V)(m_{\rm Pl}
V^\prime /V)] \nonumber\\ \label{eq:n-1}
& + & {\cal O}[(m_{\rm Pl}^3V^{(3)}/V)(m_{\rm Pl}^2
V^{\prime\prime}/V)(m_{\rm Pl}V^\prime /V)] + {\cal O}[(m_{\rm Pl}^3
V^{(3)}/V)(m_{\rm Pl}V^\prime /V)^3] +\cdots .
\end{eqnarray}

The expansion for $V$ begins with a term that involves
no derivatives; the next term involves two derivatives; the
next four derivatives, and so on.  The expansions for
$n_T$, $T/S$, and $(n-1)$ begin with terms involving two
derivatives; followed by terms involving four derivatives;
and so on.  In the previous literature, the lowest-order term has
been referred to as first-order; the next term, which involves two additional
derivatives, has been referred to as second-order; and so on.
Explicit expressions for the second-order terms
are given in Refs.~\cite{SL,second}; some of the third-order
terms for $(n-1)$ are given in Ref.~\cite{KV}.

In the next three subsections we address the convergence
of the Taylor series for the potential and the relative
sizes of the terms in the expansions for the observables.
We show that for a very general
class of potentials that lend themselves to reconstruction
that the higher-order terms in
these expansions are smaller and are bounded by
$n_T/\Delta N^{m/2-1}$, where $m$ is the number of derivatives
in the term, and further, that the Taylor series
for the potential is absolutely convergent.

Before going on, let us remind the reader of a very useful fact
and mention some notation.  The variation in the scalar field over
the eight relevant e-folds of inflation will be needed
in many places; it is $\Delta \phi /m_{{\rm Pl}} \sim (m_{{\rm Pl}} V^\prime
/V)\Delta N$, where $\Delta N \sim 8$ and throughout we
use $\Delta$ to indicate the change in a quantity over
the eight relevant e-folds.  Since $n_T \sim (m_{\rm Pl}V^\prime
/V)^2$, we will use $n_T^{1/2}$ to characterize the size
of $m_{\rm Pl}V^\prime /V$.  While it is actually
$(n-1)-3n_T$ and not $(n-1)$ whose lowest order term
is given by $m_{\rm Pl}^2V^{\prime\prime}/V$, for simplicity
we will often use $(n-1)$ to characterize the size of
$m_{\rm Pl}^2V^{\prime\prime}/V$.

\subsection{Scale-free potentials}

Let us begin with a very simple class of potentials before we consider
the general case.  These are potentials without a scale
other than an overall normalization; e.g., $V(\phi ) =
V_0\exp (-\beta \phi )$, $V(\phi ) = a\phi^b$, or
$V(\phi ) = a\phi^{-b}$.  For such potentials
there is but a single expansion parameter since
\begin{equation}\label{eq:scalefree}
{m_{{\rm Pl}}^m V^{(m)} \over V} \sim {\cal O} [(m_{{\rm Pl}} V^\prime /V)^m].
\end{equation}
For the potentials given above, Eq.~(\ref{eq:scalefree})
follows directly;  in the absence of a more quantitative definition of
scale-free we shall use Eq.~(\ref{eq:scalefree}) as the definition.

If we use $n_T\sim (m_{\rm Pl}V^\prime /V)^2$ to
characterize the deviation from scale-invariance, it follows that
\begin{equation}
{m_{{\rm Pl}}^mV^{(m)} \over V} \sim {\cal O}[n_T^{m/2}],
\end{equation}
with higher-order corrections to the expression for
$m_{{\rm Pl}}^m V^{(m)}/V$ increasing as powers
of $n_T$.  For such potentials $n-1$ and $n_T$ are
necessarily of the same order, which is not true in
the general case.  The convergence of the power
series for $V(\phi )$ over the interval of $\Delta N\sim 8$
e-folds is manifest, as the contribution of the $m$-th
derivative to the Taylor expansion is
\begin{equation}
{\Delta V_m \over V} \sim {1\over m!}\,{V^{(m)} \Delta\phi^m\over V}
\sim  \frac{(n_T \Delta N)^m}{m!}.
\end{equation}

For scale-free potentials Eq.~(\ref{eq:scalefree})
provides the ordering of terms in the
expansion of the observables in terms of the derivatives of the
potential very directly:  the order of a term involving
$m$ derivatives is $(m_{\rm Pl}V^\prime /V)^m \sim {\cal O}
(n_T^{m/2})$.  For example, in expression
Eq.~(\ref{eq:n-1}) for $(n-1)$ the first two terms are of the
order of $n_T$; the next is of the order of $n_T^2$; and the final
two are of the order of $n_T^3$.

\subsection{Strong reconstructability}

A priori we do not know the form of the potential and
thus whether or not it is scale-free; therefore,
it is important to address the most generic case possible.
Lacking a priori knowledge of the potential,
one can take advantage of the observational data itself
for guidance in reconstruction.  In the near term the observational
data available are likely to be a handful of numbers, e.g., $n_T$,
$(n-1)$, $S$, and $T$.  A reasonable, robust, and pragmatic criterion for
reconstructability is that the spectral indices do not vary
greatly over the eight e-folds of interest; that is,
\begin{equation}
{|\Delta n_T| \over |n_T|} < \delta ,\qquad {|\Delta (n-1)|\over |(n-1)|}
< \delta ,
\end{equation}
where $\delta$ is some suitably small number.
We shall refer to this as ``strong reconstructability,'' or SR.

Since the scalar and tensor spectral indices depend upon
the first two derivatives of the inflationary potential,
SR can be quantified in the following way:  $V^\prime$
and $V^{\prime\prime}$ should not vary significantly over
the interval of inflation affecting astrophysically interesting
scales.  This in turn constrains the higher derivatives of
the potential through their contributions to the Taylor expansions
of $V^\prime$ and $V^{\prime\prime}$:
\begin{eqnarray}
{\Delta V^\prime \over V^\prime} < \delta
&\Rightarrow& {V^{(m)}\Delta\phi^{m-1}\over
(m-1)! V} < \delta \,{V^\prime \over V}\qquad {\rm for\ }m\ge 2;\\
{\Delta V^{\prime\prime} \over V^{\prime\prime}}< \delta
&\Rightarrow& {V^{(m)}\Delta\phi^{m-2}
\over (m-2)! V} < \delta\, {V^{\prime\prime}\over V}\qquad {\rm for\ }m\ge3.
\end{eqnarray}
Again using $\Delta\phi /m_{{\rm Pl}} \sim (m_{{\rm Pl}} V^\prime
/V)\Delta N$, these bounds become
\begin{eqnarray}
{m_{{\rm Pl}}^mV^{(m)}\over V} & < &\left({m_{{\rm Pl}}
V^\prime\over V}\right)^{-m+2}
\Delta N^{-m+1} \delta\, (m-1)! \nonumber\\ \label{eq:ntbound}
& < & {\cal O}[n_T^{-m/2+1}\Delta N^{-m+1}\delta (m-1)!]
\qquad {\rm for\ }m\ge 2;\\
{m_{{\rm Pl}}^mV^{(m)}\over V} & < & \left({m_{{\rm Pl}}
V^\prime\over V}\right)^{-m+2}
\left({m_{{\rm Pl}}^2V^{\prime\prime}\over V}\right)\Delta N^{-m+2}
\delta\, (m-2)! \nonumber\\  \label{eq:n-1bound}
& < & {\cal O} [(n-1)n_T^{-m/2+1}\Delta N^{-m+2}\delta (m-2)!]
\qquad {\rm for\ }m\ge 3,
\end{eqnarray}
where in the final expressions we have used the fact
that $n_T \sim {\cal O}[(m_{{\rm Pl}} V^\prime /V)^2]$ and
that $(n-1)\sim {\cal O}[m_{{\rm Pl}}^2 V^{\prime\prime}/V]$.

These constraints to the derivatives of the potential are
weaker than the ones we derived for scale-free potentials, but
are more generally applicable and serve the same purpose.
The second of these implies
that the Taylor expansion for the potential is absolutely
convergent, as it
bounds the contribution to $V(\phi )$ from the $m$-th derivative
\begin{equation}
{\Delta V_m \over V} \sim \frac{V^{(m)}\Delta \phi^m}{m! V} <
{\cal O}\left( {n_T(n-1)\Delta N^2 \delta \over m(m-1)} \right).
\end{equation}

Constraints (\ref{eq:ntbound}) and (\ref{eq:n-1bound})
also serve to order terms in the expansions of the observables in
terms of the potential and its derivatives.  For
example, Eq.~(\ref{eq:n-1}) for the scalar index
includes terms of order $(m_{{\rm Pl}} V^\prime /V)^2$, $m_{{\rm
Pl}} V^{\prime\prime}/V$, $(m_{{\rm Pl}} V^\prime /V)(m_{{\rm
Pl}}^3V^{(3)}/V)$.  Our SR bounds tell us nothing about the relative
sizes of first and second terms (the first-order terms),
though they imply that the second term
must be smaller than $\delta /\Delta N$.  Based on the SR bounds, the third
term (second-order term) must be less than both $\delta
/\Delta N^2$ and $(m_{{\rm Pl}}^2V^{\prime\prime}/V) \delta /\Delta N$, and
so it is necessarily of higher order than the second term.
The higher-derivative terms,
$(m_{{\rm Pl}} V^\prime /V)^3 (m_{{\rm Pl}}^3V^{(3)}/V)$ and $(m_{{\rm
Pl}} V^\prime /V)(m_{{\rm Pl}}^2 V^{\prime\prime}/V) (m_{{\rm
Pl}}^3V^{(3)}/V)$ terms are of even higher order:
the former must be less than $\delta /\Delta N^2$ times the
first term in the expansion for $(n-1)$ and less
than $n_T \delta /\Delta N$ times
the second term in the expansion for $(n-1)$, while the latter must be less
than $\delta /\Delta N^2$ times the second term in the expansion
for $(n-1)$.  The ordering of the terms in the derivative expansion
for $(n-1)$ is clear:  more derivatives are suppressed by
powers of $\Delta N$.  In particular, a term involving $m$
derivatives can be no larger than $n_T \delta /\Delta N^{m/2-1}$.

Before ending this subsection, we mention an interesting
possibility:  for models with very large deviations from scale-invariance the 
data may some day be good enough that a small fractional change in the spectral 
index is observable.
A case in point is intermediate inflation \cite{INTER}, where the scalar index
may be greater than unity and may decrease significantly.
In particular, the potential for intermediate inflation
is scale-free and $dn /d\ln k \sim (n-1)^2$ so
that $\Delta (n-1) /(n-1) \sim \Delta N (n-1)$.
The SR bounds still apply, and additionally, the new observable $dn/d\ln k$
allows one to determine $V'''$ at lowest order (as described in section 3).

\subsection{Weak reconstructability}

The pragmatic criteria of SR discussed above can be relaxed somewhat, 
without sacrificing the convergence of the Taylor series
for the potential or the ordering of terms in the expansions
for the observables.  Suppose that one, or even both,
of the spectral indices did indeed exhibit a large fractional change over
astrophysically interesting scales, so that a
power-law description of the scale dependence of the metric
perturbations is not strictly valid.  If the absolute value of the
change is much less than unity, then the fact that $\Delta n_T > n_T$
or $\Delta (n-1) > (n-1)$ is undetectable and of little practical significance, 
and,
as we shall show now, reconstruction can proceed.  We refer to this as ``weak
reconstructability,'' or WR.
A case in point is the natural inflation
model \cite{NAT}; with parameters chosen to give $(n-1)=-0.3$,
the tensor spectral index grows by a factor of about 100 between the
largest and smallest interesting scales.  However, this growth is
entirely unobservable, being the difference between $n_T =-10^{-9}$ and
$n_T=-10^{-7}$.

Logically, there are three cases of WR:  (i) scalar index satisfies
SR and tensor index satisfies WR; (ii) tensor index satisfies
SR and scalar index satisfies WR; and (iii) both tensor and
scalar indices satisfy WR.  Since we have previously derived
the bounds to $m_{\rm Pl}^m V^{(m)}/V$ that follow if tensor
and scalar indices satisfy SR, cf. Eqs.~(\ref{eq:ntbound}) and 
(\ref{eq:n-1bound})
respectively, here we simply do the same for WR.  In case
(i) the SR scalar and WR tensor bounds apply; in case (ii) the
WR scalar and SR tensor bounds apply; and in case (iii) the scalar and tensor
WR bounds apply.  In all three cases the implications for
convergence of the power series and the ordering of terms is very similar
to the case of SR.

Let us take $\delta$ to be the parameter that quantifies the smallness
of the tensor (or scalar) index and its absolute change.
For sake of definiteness, we
would imagine that a change of a few hundredths for the scalars, and
considerably more for the tensors, would be extremely hard to observe.
Following the same strategy as in the SR case, this time
bounding the absolute change in the spectral index
due to higher-order derivatives, we find:
\begin{eqnarray}\label{eq:WRtensor}
\left({m_{{\rm Pl}}^m V^{(m)}\over V}\right) & < &
 n_T^{-m/2} \Delta N^{-m+1} \delta (m-1)!\qquad {\rm for\ }m\ge 2
{\rm \ (tensor)}, \\ \label{eq:WRscalar}
\left({m_{{\rm Pl}}^m V^{(m)}\over V}\right) & < &
 n_T^{(1-m)/2} \Delta N^{-m+2}\delta (m-2)!\qquad {\rm for\ }m\ge 3
{\rm \ (scalar)}.
\end{eqnarray}
These constraints differ from their counterparts in the
SR case only slightly:  by one fewer factor of $n_T$ (tensor) and by
the absence of the $(n-1)$ factor (scalar).  Thus, the conclusions
reached for convergence and term ordering in the SR case
carry over with only minor modification.
For example, the size of the contribution of the $m$-th
derivative to the Taylor series of the potential is
bounded by $(n-1)n_T\Delta N^2\delta  /m(m-1)$,
$n_T\Delta N^2\delta /m(m-1)$, and $n_T \Delta N^2\delta /m(m-1)$
in cases (i)--(iii) respectively; this guarantees absolute convergence.
If the SR tensor bound applies then a term involving $m$
derivatives is as before bounded by $n_T\delta /\Delta N^{m/2-1}$;
if the WR tensor bound applies then such a term is bounded
by $\delta /\Delta N^{m/2-1}$.

Finally, what types of potentials give rise to order unity fractional
changes in the spectral indices while still satisfying the
WR criteria?   It is simple to show for the tensor index that
$\Delta N (n-1)$ must be of order unity or larger; this occurs in
the previously mentioned natural inflation model.
For the scalar index the condition is that $(m_{\rm Pl}^3
V^{(m)}/V)n_T\Delta N/(n-1)$ must be of order unity or larger.

\subsection{Sensibility summary}

When physicists construct an expansion in a small
parameter (or even several small parameters) they rarely worry about
rigorous mathematical issues.
While we would like to follow in that tradition,
the problem here is a bit more vexing as there are
in principle an infinity of small expansion parameters:
$m_{\rm Pl}^mV^{(m)}/V$.
We have addressed two (not unrelated) issues here:  convergence of
the reconstructed potential and ordering of terms.

Based upon pragmatic criteria that derive from the data
themselves we have shown that convergence and term ordering
follow for potentials where the spectral indices do
not vary significantly over astrophysically interesting
scales (referred to as SR), or if they do vary by of
order unity, the absolute change is small by comparison to
what can be measured (referred to as WR).  In both cases
we explicitly showed that the Taylor expansion for the potential is necessarily
convergent, and that higher-derivative terms in the expansions for
the observables descend in size.   For ``scale-free''
potentials a term that involves $m$ derivatives is
of the order of $n_T^{m/2}$; in the more generic
cases of SR and WR, such a term is bounded by
$n_T\delta /\Delta N^{m/2-1}$
and $\delta /\Delta N^{m/2-1}$ respectively.  This establishes
what has been previously assumed implicitly in
Refs.~\cite{SL,second}:
the terms involving more derivatives are of higher order.

\subsection{The consistency relation}

An important feature of reconstruction is that the
problem is overdetermined; specifically, a set of $M\ge 3$
observables can be expressed in terms of the potential and
its first $M-2$ derivatives.  This implies a ``consistency
relation,'' which, for increasing $M$, contains terms
of higher and higher order.  The lowest-order consistency equation,
$n_T = -{1\over 7}{T\over S}$, has been much discussed (e.g., in
Ref.~\cite{expand,turner1}) and arises through Eqs.~(\ref{eqns}), (\ref{eqS}) 
and
(\ref{eqT}) which express $n_T$, $S$, and $T$ in terms of $V_{50}$ and 
$V_{50}^\prime$.

Calculating higher derivatives alone, while keeping the calculation of
each derivative to lowest order, does not lead to the correct
second-order term in the consistency equation, and nor does calculating
the second-order corrections to the derivatives present. One must
systematically do both.  The second-order version of the consistency
equation is obtained by calculating the potential, its derivative and
Eq.~(\ref{eqns}) to a higher order. Adding an extra order to the
calculation of $V_{50}'$ adds a new observable, $(n-1)$, which will
appear in the consistency equation at second order. To account for there
being still only a single consistency equation, there must be a new
equation, and because $(n-1)$ has only entered at second-order in
$V_{50}'$, we only need the first-order equation for $V_{50}''$.  The
second-order consistency equation, which we calculate in this paper,
therefore relates $n_T$, $\frac{T}{S}$ and $(n-1)$, with the last only
appearing as a second-order correction.  Were one to desire a
calculation to yet higher order, the same pattern would persist; each
existing derivative must be calculated to one extra order and the next
derivative to lowest order, introducing a new observable. This will
generate next-order terms in the consistency equation with the new
observable appearing at that order. However, this presently cannot be
done as third-order expressions for $V_{50}$ and $V_{50}'$ have not been
calculated.

\subsection{Expansion techniques}

Given the value of the potential and its first two or three derivatives
at a point and the $\phi_N$ relation just obtained, one can
reconstruct the potential on the observationally relevant scales
(i.e., $N \simeq 42-50$). The standard technique used
previously is the Taylor expansion
\begin{equation}
V(\phi) = V_{50} + V_{50}' (\phi - \phi_{50}) + \frac{1}{2} V_{50}'' 
    (\phi - \phi_{50})^2 + \cdots
\end{equation}
For many situations this is perfectly fine (e.g., when
$n_T$ and $n-1$ are small, see Ref.~\cite{turner3}).
However, if the range of eight or so e-foldings corresponds
to a large range in $\phi$ the convergence may not be
very good because of the abrupt truncation of the Taylor series.  
Specifically,
for large $(\phi - \phi_{50})$ the shape of the reconstructed
potential is dictated, rightly or wrongly, by the last term
in the expansion (quadratic or cubic).

An alternative is the Pad\'{e} approximant \cite{PTVF},
which can be generated directly from a truncated power series.
For a power series that extends to order $N$, the Pad\'{e}
approximants are quotients of two polynomials of order $L$ (numerator)
and $M$ (denominator) denoted by $[L,M]$, where $L+M=N$.  By
construction, the expansion of $[L,M]$ matches that of the
power series to order $N$, but of course is not truncated.
Very often, 
Pad\'{e} approximants provide a very good approximation
over a wider range of values than the Taylor series from
which they are derived; they in some way encode better estimates of the 
higher-order terms than does truncation. If we truncate
the Taylor series at the second derivative, then the associated
diagonal Pad\'{e} approximant $[1,1]$ is a ratio of two
first-order polynomials given by\footnote{The $[2,0]$ approximant
is just the truncated Taylor series; in addition to simplicity, there
is some motivation for using the diagonal approximant rather
than the $[0,2]$ approximant as it is asymptotically constant,
consistent with the flatness of inflationary potentials.}
\begin{equation}
R(\phi) = \frac{a_0 + a_1 (\phi - \phi_{50})}{1 + b_1 (\phi - \phi_{50})} ,
\end{equation}
with
\begin{equation}
a_0 = V_{50}; \quad b_1 = -V_{50}''/2V_{50}'; \quad
   a_1 = V_{50}' - V_{50} V_{50}''/2V_{50}' .
\end{equation}
As we shall illustrate later by specific example, Pad\'{e} approximants have 
a lot to offer when the Taylor series proves a poor approximation.

\section{Second-order Reconstruction Reduced to Practice}

Having discussed the philosophy and strategy, let us proceed to deriving
the full reconstruction equations at second-order.  The reconstruction
equations for the scalar potential and its first two derivatives,
evaluated to second-order, are given in Ref.~\cite{second}, though not in
terms of cosmological observables.  They are given in terms of the
perturbation amplitudes $A_G^2$ and $A_S^2$.  Very roughly, $A_S$ is the
horizon-crossing amplitude of the density perturbation on a given scale
and $A_G$ is the horizon-crossing amplitude of the tensor perturbation
(in the Appendix we provide some relations between notation used in that
paper and this one.)  Our purpose here is to express these second-order
expressions for the potential and its first two derivatives in terms of
the measurable quantities $n$, $d n /d\ln k$, $n_T$, $S$, and $T$.

The amplitudes $A_S^2$ and $A_G^2$
are related to the observables $S$, $T$, $n_T$ and $n$ by:
\begin{equation} 
\label{rosetta} 
A_G^2 = 0.70(1 - 1.3n_T)T, \qquad A_S^2 =
9.6[1 - 1.15(n-1)]S, 
\end{equation} 
where the $(n-1)$ and $n_T$ dependencies
have been found by evaluating the Sachs-Wolfe integrals numerically.  Both
expressions are accurate to second-order.

Before deriving second-order expressions for the potential and its
derivatives, we calculate the second-order version of the consistency
relation.  It is obtained from Eq.~(2.9) of Ref.~\cite{second},
\begin{equation}
-{n_T\over 2} = {A_G^2\over A_S^2}\, \left[ 1+
3\epsilon -2\eta\right], 
\end{equation} 
where to the required order the slow-roll
parameters $\epsilon$ and $\eta$ (defined in the Appendix) are given by
\begin{equation} 
\epsilon = -n_T/2, \qquad \eta = (n-1)/2 - n_T.
\end{equation} 
This gives a simple and very useful relation for $A_G^2/A_S^2$,
\begin{equation} 
{A_G^2\over A_S^2} = -0.5n_T \left[1-0.5n_T +1.0(n-1)\right] .
\end{equation} 
Substituting into Eq.~(\ref{rosetta}), we find the second-order consistency 
relation
\begin{equation}
n_T= -{1\over 7}{T\over S} \left[ 1 - 0.8 n_T + 0.15 (n-1)\right],
\end{equation} 
or 
\begin{equation} 
\label{eq:consis} 
{T\over S} = -7 n_T \left[ 1 +0.8 n_T - 0.15 (n-1) \right].
\end{equation}
To the required order we can use the first-order truncation $n_T = -{1\over
7}{T\over S}$ inside the brackets, thereby obtaining an alternative form,
\begin{equation}
\label{eq:consis2} 
n_T = -{1\over 7}{T\over S}\left[ 1 + 0.11 {T\over S} + 0.15 (n-1) \right] ,
\end{equation} 
where $n_T$ is given in terms of the more
accessible quantities $(n-1)$ and ${T\over S}$.

Independent measurements of $n$, $n_T$ and $T\over S$
provide a powerful test of the inflationary hypothesis; in the space of
these parameters inflationary models
must lie on the surface defined by Eq.~(\ref{eq:consis}).
In Figure 1 we illustrate the inflationary surface both without and
with second-order corrections.  The second-order corrections break the
degeneracy in the $(n-1)$ direction, as well as typically
reducing $T\over S$ viewed
as a function of $n_T$ and $(n-1)$.  However,
the portions of the surface that feature large
corrections are not favored by present cosmological data,
and further, are susceptible to higher-order corrections. (Indeed, well away 
from scale-invariance the surface would be noticeably different even just 
using
Eq.~(\ref{eq:consis2}) instead of Eq.~(\ref{eq:consis}), which
differ by third and higher order terms.)

Obtaining the reconstruction equations is simply a matter of substituting
into Eqs.  (3.4), (3.6) and (3.15) of Ref.~\cite{second} for $V$, $V'$ and
$V''$ respectively.  We give two alternative forms for each, the first using
$n_T$ and the second substituting $T\over S$ for $n_T$ using the second-order
consistency equation.  They are 
\begin{eqnarray} \label{eq:second}
V_{50}/m_{{\rm Pl}}^4 &
= & 1.65(1-1.4 n_T)T,\nonumber\\
& = & 1.65\left( 1 + 0.20{T\over S} \right) T,\\
V_{50}'/m_{{\rm Pl}}^3 & = & \pm 8.3\sqrt{-n_T} \, [1
- 1.1 n_T - 0.03 (n-1)] T, \nonumber\\
& = & \pm 8.3 \sqrt{{1\over 7}{T\over S}}
\left[ 1 +0.21 {T\over S} -0.04(n-1) \right] T, \\
V_{50}''/m_{{\rm Pl}}^2 & = & 21 \Biggl[ (n-1) - 3 n_T + 1.4
n_T^2 \nonumber\\ 
&\qquad & + 0.6 n_T(n-1) -0.2 (n-1)^2 + 1.1{dn\over d\ln
k} \Biggr] T, \nonumber\\
& = & 21 \Biggl[ (n-1) +0.43{T\over S} +0.073 \left(
{T\over S}\right)^2 \nonumber\\ 
&\qquad & -0.015{T\over S}(n-1) -0.2(n-1)^2 +
1.1{dn\over d\ln k} \Biggr] T .
\end{eqnarray} 
These expressions are accurate to second-order.
Naturally, they agree with the first-order expressions given
earlier. 

Though no expression is given in Ref.~\cite{second} for $V^{\prime \prime
\prime}$, by using the lowest-order expressions for $\epsilon$,
$\eta$, and a third slow-roll parameter $\xi$, and Eq.  (3.13) which relates 
the three to
$dn/d\ln k$, one can obtain the first-order expression,
\begin{eqnarray} 
V_{50}'''/m_{{\rm Pl}} & = & \pm
104\sqrt{-n_T}\left[ {dn/d\ln k \over n_T}
- 6n_T + 4(n-1) \right]T , \nonumber\\
& = & \pm 104 \sqrt{{1\over 7}{T\over S}} \left[ -7 {dn/d\ln k\over T/S}
+ 0.9 {T\over S} + 4(n-1) \right]T.
\end{eqnarray} 
where the overall sign is to be the same as that of $V'$.
The second-order term would require yet another observable.  As
remarked in Ref.~\cite{second}, even this first-order expression
features the rate of change of the scalar spectral index,
which is likely to be very difficult
to measure.  Realistically then, in the near term only the value of the
potential and its first two derivatives are likely to
be accessible to accurate  determination.  

The final step in reconstructing the potential is to use $d\phi /dN$ to
the desired order, to find the range of $\phi$ that
corresponds to the eight or so e-foldings of inflation relevant for
astrophysics.  To proceed, we may simply carry out a Taylor expansion
of $\phi$ about $\phi_{50}$, to whatever order we believe is appropriate,
\begin{equation} 
\phi_N - \phi_{50} = \left.
(N-50) \frac{d\phi}{dN}\right|_{\phi_{50}}+\  \frac{1}{2}(N-50)^2 \left.
\frac{d^2\phi}{dN^2}\right|_{\phi_{50}} + \cdots
\end{equation}
This is a double expansion, in the sense that the coefficients are
themselves obtained as a series expansion in the slow-roll parameters.

To proceed, we use as a starting point the {\em exact} formula
\begin{equation} 
\dot{\phi} = - \frac{m_{{\rm Pl}}^2}{4\pi} H' \,,
\end{equation} 
which, along with $dN/dt = -H$, yields the relation from which
the Taylor coefficients may be calculated,
\begin{equation} 
\frac{d\phi}{dN} = \frac{m_{{\rm Pl}}^2}{4\pi} \frac{H'}{H} .
\end{equation} 

To get a given coefficient in the Taylor expansion for $\phi_N$,
one simply calculates $d^i\phi /dN^i$ expanding to the desired
order in the deviation from scale invariance.\footnote{Note this procedure
differs slightly from that in Ref.~\cite{turner3}, where $d\phi/dN$
was expanded linearly about $\phi_{50}$ and
$\phi_N$ was solved for exactly, cf. Eq.~(8).  This results in an
exponential, whose expansion picks up the $(N-50)$ and $(N-50)^2$ terms
correctly to lowest order in the deviation from scale invariance,
though not the higher-order terms in
the $(N-50)$ term which would require higher-order terms in the
expansion of $d\phi/dN$.  There is an overall sign error in
Eq.~(8) of Ref.~\cite{turner3}.}  For example, taking only the first
term in the $\phi_N$ expansion and working to first-order yields the
expression already given in Section 1.1.
We give the first coefficient in the $\phi_N$ expansion
to second-order and the second coefficient in the $\phi_N$
expansion to first-order only,
\begin{eqnarray}\label{eq:delphi}
\phi_N -
\phi_{50} & = & \pm \frac{m_{{\rm Pl}}}{\sqrt{8\pi}} \sqrt{-n_T} \, \left[ 1
+ 0.1 n_T + 0.1 (n-1) \right] (N-50) \nonumber\\
& & \pm \frac{m_{{\rm Pl}}}{4\sqrt{8\pi}} \sqrt{-n_T} \,
\left[ (n-1)-n_T \right] (N-50)^2 + \cdots ,
\end{eqnarray} 
with both signs again agreeing with that of $V'$.

In the process of reconstruction, we shall use the
first-order expansion for $\phi_N -\phi_{50}$ in
first-order reconstruction, and the second-order
expansion in second-order reconstruction.

\section{Reconstructing an exponential potential}

A useful testing ground for reconstruction is the exponential
potential, the only known case where the perturbation spectra can be
derived exactly analytically \cite{SL,LS}.  For the potential
\begin{equation}
V(\phi )=V_0\exp\left( -\sqrt{\frac{16\pi}{p}}\,\frac{\phi}{m_{Pl}} \right) ,
\end{equation} 
the scale factor grows exactly as $t^p$.  Compared with the lowest order
expressions, the amplitudes $A$ and $A_T$, or $A_S^2$ and $A_G^2$,
are both multiplied by the same $p$-dependent factor $R^2(p)$, where
\begin{equation}
R(p) = 2^{1/(p-1)} \frac{\Gamma\left[3/2 + 1/(p-1)\right]}{\Gamma[3/2]}
\left(1-1/p \right)^{p/(p-1)} ,
\end{equation} 
where $\Gamma (\cdots )$ is the usual gamma function.
Both scalar and tensor spectra are exact
power laws with spectral indices $(n-1) = n_T = -2/(p-1)$.
The scalar-field solution is characterized by
\begin{equation}
\dot\phi = \sqrt{p\over 4\pi} \,{m_{\rm Pl}\over t};\quad
{d\phi \over dN} = -{m_{\rm Pl}\over \sqrt{4\pi p}};\quad
V(\phi_N) = V_{50}^{{\rm true}} \exp \left[ 2(N-50)/p\right] .
\end{equation}
The expressions for $T$ and $S$ can be obtained exactly
by integrating Eqs.~(\ref{eq:sws}) and (\ref{eq:swt}),
\begin{eqnarray}\label{eq:S}
S & = & 2.2 f(n) R^2(p) {V_{50}^{{\rm true}} \over m_{\rm Pl}^4x_{50}^2} ,\\
\label{eq:T}T & = & 0.61 g(n_T) R^2(p) {V_{50}^{{\rm true}} 
  \over m_{\rm Pl}^4},
\end{eqnarray}
where the numerical factors $f(n) = 1 +1.15(n-1) +\cdots$
and $g(n_T) = 1 + 1.3n_T +\cdots$ arise from the
$n$, $n_T$ dependence of the Sachs-Wolfe integrals, cf. Eqs.~(\ref{eq:sws},
\ref{eq:swt}).

We are now ready to carry out an array of reconstruction methods.
Because we are using exact expressions to generate the spectra, this
procedure is more ambitious, and more realistic, than those attempted
thus far \cite{copeland,turner3}, where the trial spectra were produced
using the slow-roll approximation.  For the general inflationary
potential, exact results are not known, and so this procedure is not
possible\footnote{Of course, we are going to pretend that we don't know
the potential is exponential to demonstrate our methods.}.  However, our
method here should give a more realistic estimate of inherent errors
even in the general case.

There are two distinct types of error. The first is error in
the value of the potential at $\phi_{50}$, due
to third-order and higher terms.  By substituting the expression for
$T$ in Eq.~(\ref{eq:T}) into Eq.~(\ref{eq:fv}) or (\ref{eq:second})
for $V_{50}$ we can compute that error:
\begin{equation}\label{eq:under}
V_{50}/V_{50}^{{\rm true}} = g(n_T) (1-1.4n_T)R(p)^2 .
\end{equation}
The second error involves the shape of the potential,
which depends on the ability of the chosen expansion to match the
potential over the eight interesting e-foldings.

We have chosen as a specific example an exponential potential with 
$p=43/3$.   We did so because
this leads to about the largest departure from
scale invariance that can still be regarded as observationally
viable, $(n-1)=n_T =-0.15$ and ${T\over S}\simeq 1$,
and thus realistically represents the most
challenging example of reconstruction.  The exact potential is
shown in Fig.~2 along with the results of five different reconstructions.

To begin, consider the error in estimating $V_{50}^{{\rm true}}$; we have
$g(n_T=-0.15) = 0.824$ and so $V_{50}/V_{50}^{{\rm true}} \simeq 0.95$,
a modest 5\% error due to the neglected higher-order terms.  As we always
include the second-order term in $V_{50}$, the error is the same in
every method we look at. Had the first-order expression for $V_{50}$
been used instead, corresponding to the neglect of the factor of
$(1-1.4n_T)$ in Eq.~(\ref{eq:fv}), then the underestimation would have
been about 20\%.

Let us now consider the shape, which we note depends on $T$ and $S$ only 
through their ratio. The important distinction between
different methods is the difference in required input data;
methods needing only $n$ and $T\over S$ have the
advantage of depending only on the information that is easiest to obtain. 
Requiring $dn/d\ln k$ in addition, while offering more accuracy, is
setting a much trickier observational task, though upper bounds are also useful 
in the absence of actual determinations.

As a starting point, let us take the equations derived in Ref.~\cite{turner3},
which are primarily first-order though they include the second-order 
correction to $V_{50}$, cf. Eqs.~(\ref{eq:fv}--\ref{eq:fvpp}).  In this 
extreme example, the quadratic Taylor series based upon this does a bad
job of approximating the shape of the potential,
as it turns upward for large $(\phi -\phi_{50})$
due to the truncation at the $(\phi -\phi_{50})^2$ term (see Fig.~2).

If we now require knowledge of $dn/d\ln k$, the
Taylor series approach can  be improved in two ways.
We can now take $V_{50}$, $V_{50}'$, and $V_{50}''$ to second-order; however,
the improvement is rather minimal.   Alternatively,
we can stick to first-order expressions, but include the
$V_{50}'''$ cubic term.  Again the improvement is modest, though
at least the unwanted minimum has been eliminated.
One could go further and take $V_{50}$, $V_{50}'$, and $V_{50}''$ to
second-order and $V_{50}'''$ to lowest order, which we haven't
illustrated, again seeing only modest gains for
the increased observational requirement.

The Taylor series having been unimpressive, let us progress in a
different direction.  With only $n$ and $T\over S$, as an alternative to
the Taylor series one can construct the Pad\'{e} approximant based upon
it, taking $V_{50}$ to second-order and $V_{50}'$ and $V_{50}''$ to
first-order.  This represents a substantial gain on the Taylor series to
that order {\em without requiring any additional input information}.
With this minimal information, it is a much better method.  Reintroducing
$dn/d\ln k$ allows this method to be extended to second-order,
where the reproduction of the shape of the potential is excellent.  To
include the third derivative term would necessitate a more complicated
(non-diagonal) Pad\'{e} approximant, which doesn't seem warranted at the
moment.

What is the upshot of this comparison?  Recalling that we have chosen an
example with extreme deviation from scale-invariance, the second-order
corrections are reassuringly small and only improve the shape of the
reconstructed potential slightly.  The addition of the third derivative
term in the Taylor series gives a slightly more significant improvement,
but at the price of its dependence upon $dn/d\ln k$ even at lowest
order. The most remarkable improvement involves the use of Pad\'{e}
approximants.  Even without knowledge of $dn/d\ln k$ the shape of
the potential is reproduced far better than with the higher-order Taylor
series which does require that knowledge.  As noted previously, the
improvement results from the fact that the Pad\'{e} approximant is not
truncated; further, even in situations where truncation of the Taylor
series does not lead to problems, the Pad\'{e} approximant still proves
valuable as its Taylor expansion coincides with that of the original
expansion.  We therefore conclude that Pad\'{e} approximants provide a
significant improvement in the perturbative reconstruction of the
inflationary potential.

\section{Discussion}

By presenting the second-order reconstruction equations
directly in terms of observables, we have been able to
assemble and to compare an array of different perturbative
reconstruction techniques based upon cosmological observables.
Our work extends previous work in several important ways.

First, we have placed the perturbative reconstruction
process on a firmer foundation by addressing the important
issues of convergence and term ordering.
We have emphasized that the observational data themselves can
be used to decide whether or not perturbative reconstruction
is well justified and sensible.  In particular, we have
shown that the Taylor series for the potential is absolutely
convergent and that terms in the expansions for the observables
must decrease in size as the number of derivatives increase
for the case where the spectral indices do not vary
significantly over the astrophysically interesting scales, or,
if they do, their absolute change is small.

Perhaps our most interesting result is the introduction of the Pad\'{e}
approximant as an alternative to the Taylor series in perturbative
reconstruction.  It can be obtained from a Taylor series regardless of
the order (in the deviation from scale invariance) to which the
coefficients of the Taylor series has been obtained.  In our worked
example, the improvement in reproducing the shape of the potential as
compared to the Taylor series is striking, especially considering that no
extra observables are required.

We have shown that the second-order corrections to the Taylor series
coefficients are generally small, and that those for $V_{50}$ and
$V_{50}'$ only depend upon the same quantities as the first-order
expressions ($S$, $T$, and $n$). The corrections to $V_{50}''$ however
require a new observable such as $dn/d\ln k$, and by deriving for
the first time an explicit expression we have confirmed that even the
lowest-order term in $V_{50}'''$ requires this challenging observable.

Finally, one of the most important aspects of reconstruction is that it
is overdetermined: Any set of cosmological observables supplies
degenerate information regarding the potential and its derivatives,
thereby providing an important consistency check.  In particular, the
tensor spectral index can be expressed to second-order in terms of $S$,
$T$, and $n$ by the relation: $n_T = -{1\over 7}{T\over S} [1 +
0.11{T\over S} + 0.15(n-1)]$.  In cases that are observationally viable,
the second-order corrections are small.

\section*{Acknowledgements} 

We are grateful to John Barrow for bringing Pad\'{e}
approximants to our attention and to Sharon Vadas for many
stimulating discussions.   We also thank Ed Copeland, Scott Dodelson,
Rocky Kolb, and Jim Lidsey.  ARL was
supported by the Royal Society, and acknowledges the hospitality of
Fermilab where this work was initiated and also the use of the Starlink
computer system at the University of Sussex.  MST was supported in part
by the Department of Energy (at Chicago and Fermilab) and by the NASA
through grant NAGW-2381 (at Fermilab). ARL and MST both acknowledge the
hospitality of the CfPA, Berkeley, while some of this work was carried out.

\section*{Appendix:  Some Relations between Notation}

For the convenience of the reader, we summarize here some relations
between the notation used here and that in Ref.~\cite{second}, from which
several important results were taken.  In that paper, the spectra $A_S$
and $A_G$ were defined so as to include any scale-dependence within them,
i.e., they are functions of $k$.  In circumstances where the spectra can
be approximated by power-laws, these are related to the amplitudes $A$
and $A_T$ in this paper, which are just numbers, by
\begin{eqnarray} 
A (k/k_{50})^{n-1} & = & \frac{2\pi^2}{H_0^4} A_S^2(k) ,\\ 
A_T (k/k_{50})^{n_T} & = & 2 A_G^2(k) .
\end{eqnarray} 
Even in cases where the spectra cannot be described by power-laws, the
correspondence holds at $k = k_{50}$.

In Ref.~\cite{second}, slow-roll parameters $\epsilon$
and $\eta$ are introduced,
\begin{equation} 
\epsilon = \frac{m_{Pl}^2}{4\pi} \left( \frac{H'}{H}
\right)^2 , \qquad \eta = \frac{m_{Pl}^2}{4\pi} \frac{H''}{H},
\end{equation} 
which are again in general $k$-dependent.  As indicated in
Section 2 of the present paper, they can be related to the spectral indices
to various orders, $\epsilon$ and $\eta$ being of the same order in
perturbation theory as $(n-1)$ and $n_T$.  To lowest-order they are constant,
corresponding to power-law spectra.  At lowest-order $\epsilon = 16 \pi x^2$,
but higher order corrections break this relation.

\section*{Figure Captions} \bigskip

\noindent{\bf Figure 1:}  The consistency plane for inflation in 
$n-n_T-{T\over S}$ space, the flat surface being the lowest-order result
and the curved one incorporating the second-order corrections, given by
Eq.~(\ref{eq:consis}).

\medskip 
\noindent{\bf Figure 2:}  An array of different reconstructions of an 
exponential potential with $(n-1)=n_T=-0.15$ ($p = 43/3$).  The longer
dotted line indicates the exact potential.  The three different line
styles correspond to three different reconstruction strategies; solid
is Taylor series truncated at $(\phi-\phi_{50})^2$, dashed is Taylor
series truncated at $(\phi-\phi_{50})^3$ and dash-dotted is the
Pad\'{e} approximant based on the former of these. The upper line of a
given style uses coefficients to first-order in the deviation from
scale invariance (save $V_{50}$, which is always second-order), while
the lower, where plotted, is second-order in all coefficients.  The
length of the curves corresponds to eight e-foldings.

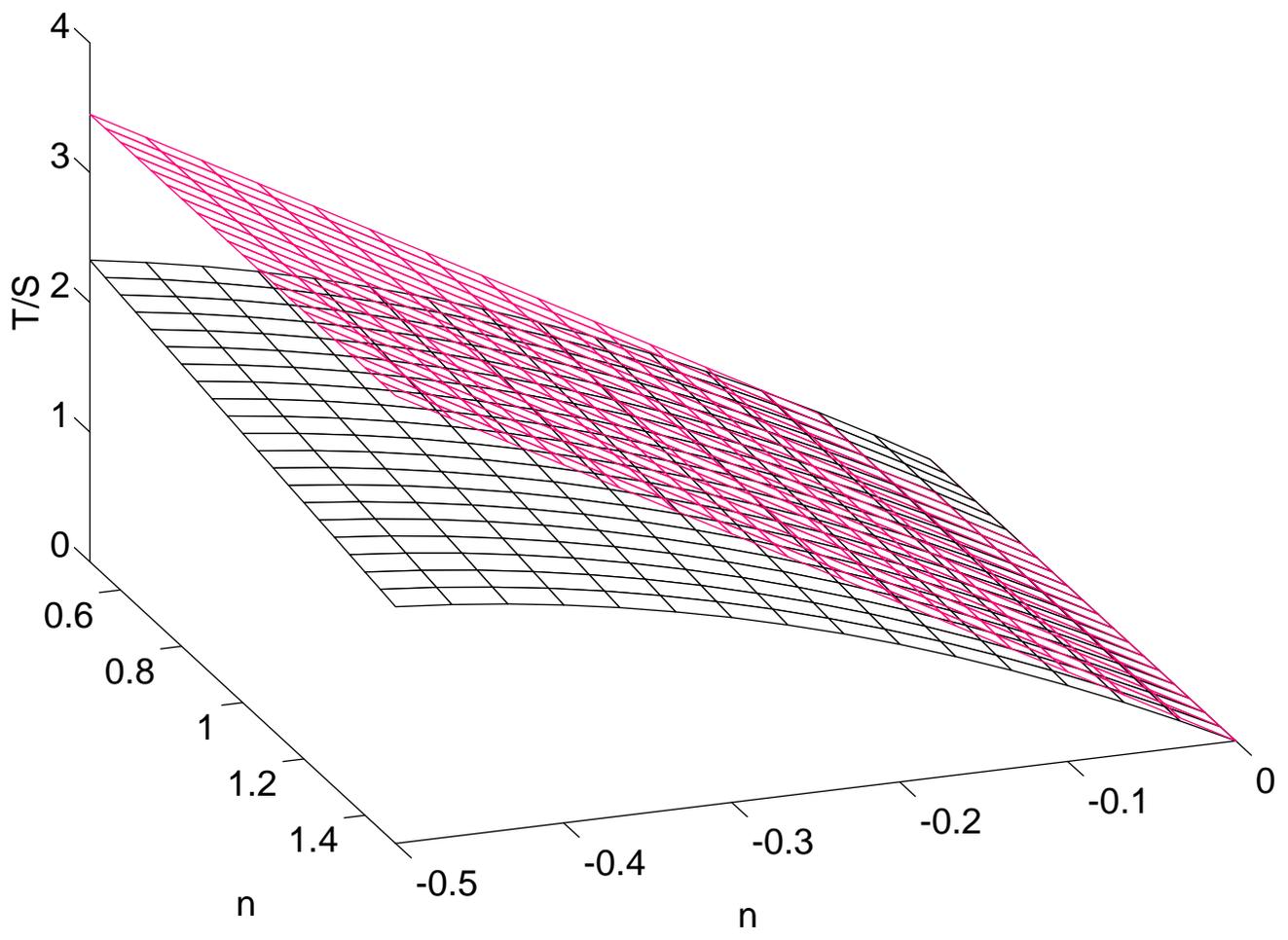
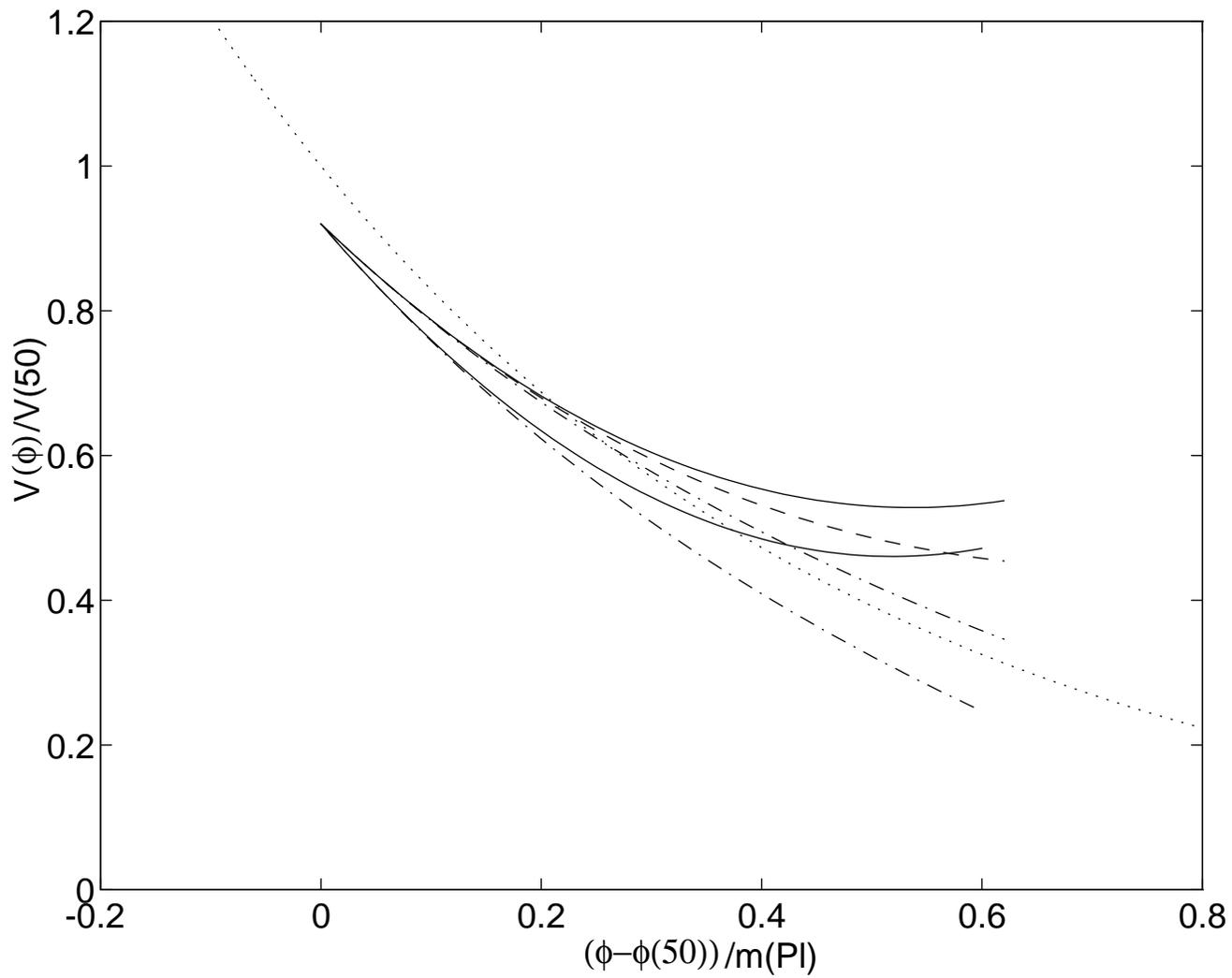
\end{document}